\documentclass[11pt, a4paper]{article}
\pdfoutput=1

\usepackage[height=8.85in,width=6.75in]{geometry}

\usepackage{graphicx,rotating}     
\usepackage[bookmarksopen,colorlinks=true,linkcolor=steelblue,
citecolor=darkred,urlcolor=darkred,linktoc=all]{hyperref}

\usepackage{amsmath,amssymb}
\usepackage{mathtools}
\usepackage{slashed}
\usepackage{bm}
\usepackage{bbm}
\usepackage{cite}
\usepackage{comment}
\usepackage[utf8]{inputenc}
\usepackage{accents}
\usepackage[footnotesize]{caption}
\usepackage{chngcntr}
\counterwithin{equation}{section}
\usepackage{cancel}
\usepackage{physics}
\usepackage{here}

\usepackage[lf]{Baskervaldx} 
\usepackage[bigdelims,vvarbb]{newtxmath} 
\usepackage[cal=boondoxo]{mathalfa} 

\usepackage{stmaryrd}
\usepackage{trimclip}
\usepackage[framemethod=default]{mdframed}
\newmdenv[skipabove=7pt,
skipbelow=7pt,
rightline=false,
leftline=false,
topline=false,
bottomline=false,
backgroundcolor=gray!10,
linecolor=gray,
innerleftmargin=5pt,
innerrightmargin=5pt,
innertopmargin=5pt,
innerbottommargin=5pt,
leftmargin=0cm,
rightmargin=0cm,
linewidth=4pt]{eBox}
\newmdenv[skipabove=7pt,
skipbelow=7pt,
rightline=true,
leftline=true,
topline=true,
bottomline=true,
backgroundcolor=white,
linecolor=gray,
innerleftmargin=5pt,
innerrightmargin=5pt,
innertopmargin=5pt,
innerbottommargin=5pt,
leftmargin=0cm,
rightmargin=0cm,
linewidth=1pt]{eBox2}

\usepackage{xcolor}
\definecolor{darkred}{rgb}{0.7, 0., 0.}
\definecolor{orangered}{rgb}{1,0.27,0.}
\definecolor{steelblue}{rgb}{0.275,0.51, 0.706}
\definecolor{forestgreen}{rgb}{0.13,0.55,0.13}
\newcommand{\Mpl}{M_{\text{Pl}}}

\usepackage{tikz}
\usepackage{tikz-feynman}
\tikzfeynmanset{compat=1.0.0}
\pgfdeclarelayer{bg}    
\pgfsetlayers{bg,main}  
\usetikzlibrary{decorations}

\pgfdeclaredecoration{dashsoliddouble}{initial}{
  \state{initial}[width=\pgfdecoratedinputsegmentlength]{
    \pgfmathsetlengthmacro\lw{.3pt+.5\pgflinewidth}
    \begin{pgfscope}
      \pgfpathmoveto{\pgfpoint{0pt}{\lw}}%
      \pgfpathlineto{\pgfpoint{\pgfdecoratedinputsegmentlength}{\lw}}%
      \pgfmathtruncatemacro\dashnum{%
        round((\pgfdecoratedinputsegmentlength-3pt)/6pt)
      }
      \pgfmathsetmacro\dashscale{%
        \pgfdecoratedinputsegmentlength/(\dashnum*6pt + 3pt)
      }
      \pgfmathsetlengthmacro\dashunit{3pt*\dashscale}
      \pgfsetdash{{\dashunit}{\dashunit}}{0pt}
      \pgfusepath{stroke}
      \pgfsetdash{}{0pt}
      \pgfpathmoveto{\pgfpoint{0pt}{-\lw}}%
      \pgfpathlineto{\pgfpoint{\pgfdecoratedinputsegmentlength}{-\lw}}%
      \pgfusepath{stroke}
    \end{pgfscope}
  }
}

\begin{document}

\hypersetup{pageanchor=false}
\begin{titlepage}

\begin{center}

\hfill KEK-TH-2692

\vskip 1in

{\Huge \bfseries
Reheating after Axion Inflation
} \\
\vskip .8in

{\Large Tomohiro Fujita$^{\lozenge}$, Kyohei Mukaida$^{\blacklozenge}$, Tenta Tsuji$^{\blacklozenge}$}

\vskip .3in
\begin{tabular}{ll}
$^{\lozenge}$ & \!\!\!\!\!\emph{Department of Physics, Ochanomizu University, 2-2-1 Otsuka, Tokyo 112-8610, Japan}\\
$^{\lozenge}$ & \!\!\!\!\!\emph{RESCEU, The University of Tokyo, 7-3 Hongo, Bunkyo, Tokyo 113-0033, Japan}\\
$^{\lozenge}$ & \!\!\!\!\!\emph{Kavli IPMU (WPI),
UTIAS,
The University of Tokyo, Chiba 277-8583, Japan}\\
$^{\blacklozenge}$ & \!\!\!\!\!\emph{Theory Center, IPNS, KEK, 1-1 Oho, Tsukuba, Ibaraki 305-0801, Japan}\\
$^{\blacklozenge}$ & \!\!\!\!\!\emph{Graduate University for Advanced Studies (Sokendai), }\\[-.15em]
& \!\!\!\!\!\emph{1-1 Oho, Tsukuba, Ibaraki 305-0801, Japan}
\end{tabular}

\end{center}
\vskip .6in

\begin{abstract}
\noindent
We investigate the reheating process in an axion inflation model where the inflaton couples to non-Abelian gauge fields via the Chern--Simons coupling.
The Chern--Simons coupling leads to the efficient production of gauge fields via a tachyonic instability during inflation, whose implications have been actively studied in the literatures.
Moreover, it has been recently pointed out that the produced gauge fields can be even thermalized during inflation, leading to warm inflation.
Apparently, these findings seem to imply that the reheating is completed immediately after inflation because the tachyonic instability or the thermal friction induced by the Chern--Simons coupling cause the inflaton condensate to decay rapidly.
Contrary to this naive expectation, however, we show that, in most of the parameter space, either the inflaton condensate, the inflaton particles, or the glueballs once dominate the Universe and their perturbative decay completes the reheating.
\end{abstract}

\end{titlepage}

\tableofcontents
\renewcommand{\thepage}{\arabic{page}}
\renewcommand{\thefootnote}{$\natural$\arabic{footnote}}
\setcounter{footnote}{0}
\hypersetup{pageanchor=true}

\newpage

\section{Introduction}
\label{sec:introduction}

Inflation has been widely accepted as the standard paradigm of primordial cosmology, as it naturally explains the flatness, isotropy, and homogeneity of the observed universe, while successfully accounting for the generation of primordial fluctuations imprinted in the cosmic microwave background and large-scale structure~\cite{Weinberg:2008zzc,Baumann:2022mni}. However, the underlying microphysical mechanism responsible for inflation remains an open question. Inflation occurs when a scalar field, known as the inflaton, slowly rolls down its potential. Achieving slow-roll for a sufficiently long time has been an outstanding problem~\cite{Copeland:1994vg, Stewart:1996ey}.

One of the well-motivated approaches to address this issue is to identify the inflaton as a pseudo Nambu--Goldstone boson, namely \textit{axion}, that enjoys a weakly broken shift symmetry controlling a potential pitfall of radiative corrections.
The original idea based on this framework is called natural inflation~\cite{Freese:1990rb, Kim:2004rp, Pajer:2013fsa}, and has been extended to various axion inflation models including attempts~\cite{Croon:2014dma,Czerny:2014wza,Landete:2017amp,Nomura:2017ehb,DAmico:2021fhz} to match the current observations~\cite{Planck:2018jri,BICEP:2021xfz}.
In addition, the axion inflation naturally embraces the coupling to radiation responsible for reheating without spoiling the flat potential, such as the Chern--Simons (CS) coupling to the gauge fields.
The inflation and subsequent reheating by the CS coupling is particularly interesting as the tachyonic instability drives the rapid gauge field production~\cite{Turner:1987bw,Garretson:1992vt,Anber:2006xt,Caprini:2014mja,Adshead:2016iae}, leading to various phenomenological implications~\cite{Anber:2015yca,Domcke:2019mnd,Domcke:2022kfs,Turner:1987bw,Garretson:1992vt,Anber:2006xt,Caprini:2014mja,Fujita:2015iga,Adshead:2016iae,Gorbar:2021zlr,Domcke:2018eki,Domcke:2018gfr,Domcke:2019qmm,Gorbar:2021rlt,Fujita:2022fwc,vonEckardstein:2024tix,Cook:2011hg,Barnaby:2011qe,Anber:2012du,Namba:2015gja,Domcke:2016bkh,Adshead:2019lbr,Caravano:2022epk}.

Another intriguing approach is warm inflation, where the inflaton field is coupled to a thermal bath of radiation, leading to a friction term that sustains the slow-roll condition~\cite{Berera:1995ie,Berera:1995wh}.
This idea has now been drawing renewed attention as a possible way out to address the swampland conjecture~\cite{Motaharfar:2018zyb}.
Moreover, the warm inflation opens up a possibility that the radiation forming the thermal bath becomes dominant immediately after inflation, which realizes a maximal temperature of the Universe for a given inflaton potential~\cite{Kamali:2023lzq}.
However, the warm inflation generically suffers from the large corrections due to the thermal potential~\cite{Yokoyama:1998ju} as a counterpart of required thermal friction, spoiling the slow-roll condition.

Recently, models that combine these two ideas have been actively studied.
The shift symmetric nature of axion-like inflaton suppresses the thermal potential while it simultaneously allows the friction term, providing a natural way to realize the warm inflation~\cite{Visinelli:2011jy, Ferreira:2017lnd, Kamali:2019ppi,Berghaus:2019whh}.
A particularly promising direction involves non-Abelian gauge fields, which can rapidly reach thermal equilibrium via self-interactions~\cite{Kamali:2019ppi,Berghaus:2019whh,Das:2019acf,DeRocco:2021rzv,Laine:2021ego, Montefalcone:2022jfw,Mukuno:2024yoa}.\footnote{
  In contrast to $\mathrm{U}(1)$ gauge theory, the non-Abelian gauge theory breaks both electric and magnetic one-form symmetries. This also enhances the rapid thermalization since the color electric/magnetic fluxes are screened immediately.
}
Surprisingly, the warm inflation is \textit{not an option} of axion inflation rather an \textit{automatic outcome} for some parameters~\cite{Ferreira:2017lnd,DeRocco:2021rzv}.
Even if the gauge fields are absent initially, they are efficiently produced during inflation via the tachyonic instability, leading to the thermal bath formation.
This effect is expected to modify the predictions of axion inflation, such as the scalar non-Gaussianity, which could be tested by future observations~\cite{Moss:2011qc, Bastero-Gil:2014raa}.
In this work, we investigate the reheating process in a model where an axion-like inflaton is coupled to a non-Abelian gauge field via the CS coupling.
The lattice simulations in $\mathrm{U}(1)$ gauge theories indicate that instabilities cause the inflaton condensate to decay rapidly after inflation~\cite{Adshead:2015pva,Cuissa:2018oiw,Figueroa:2023oxc,Figueroa:2024rkr}, while warm inflation models suggest that thermal friction quickly dissipates the condensate.
At first glance, these observations seem to imply that reheating in this model is also completed immediately after inflation.
However, as we will demonstrate, neither scenario adequately describes our model.
In fact, we show in detail that, over most of the parameter space, reheating is completed via the perturbative decay of either the inflaton condensate, the inflaton particles, or the glueballs.

\begin{figure}[H]
	\centering
  \includegraphics[width=0.7\linewidth]{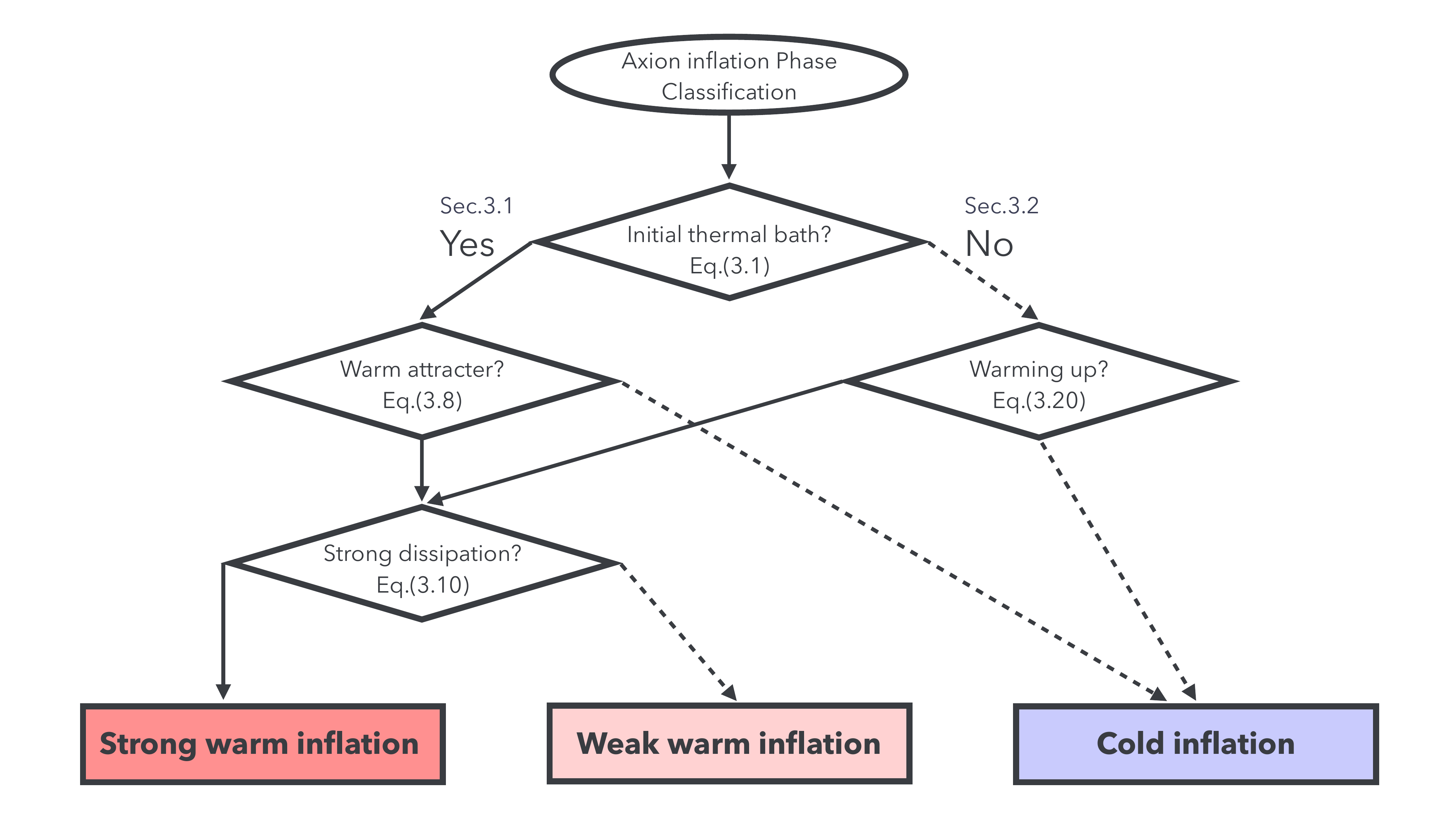}
  \hfill
  \includegraphics[width=0.7\linewidth]{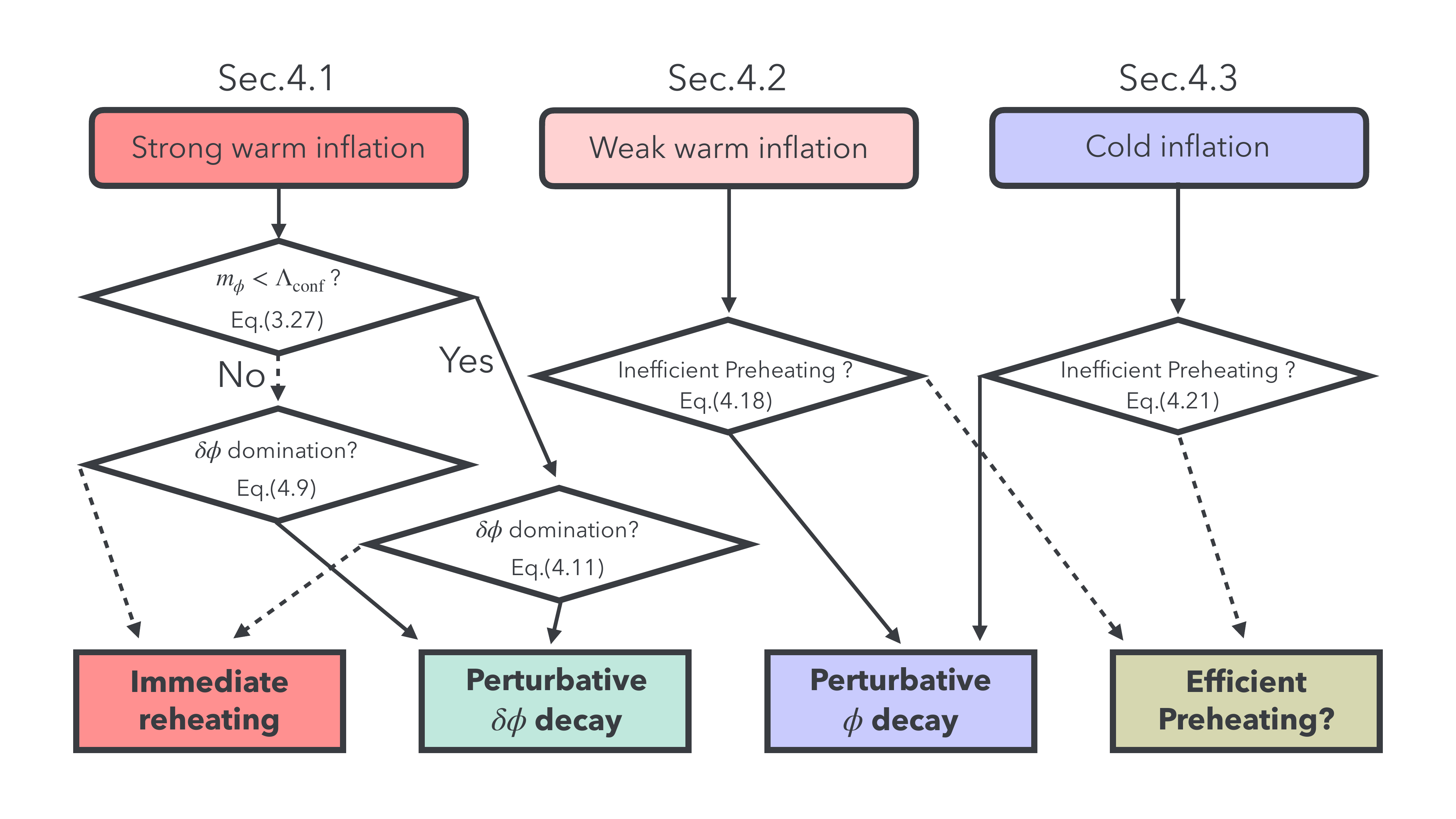}
  \hfill
  \includegraphics[width=0.7\linewidth]{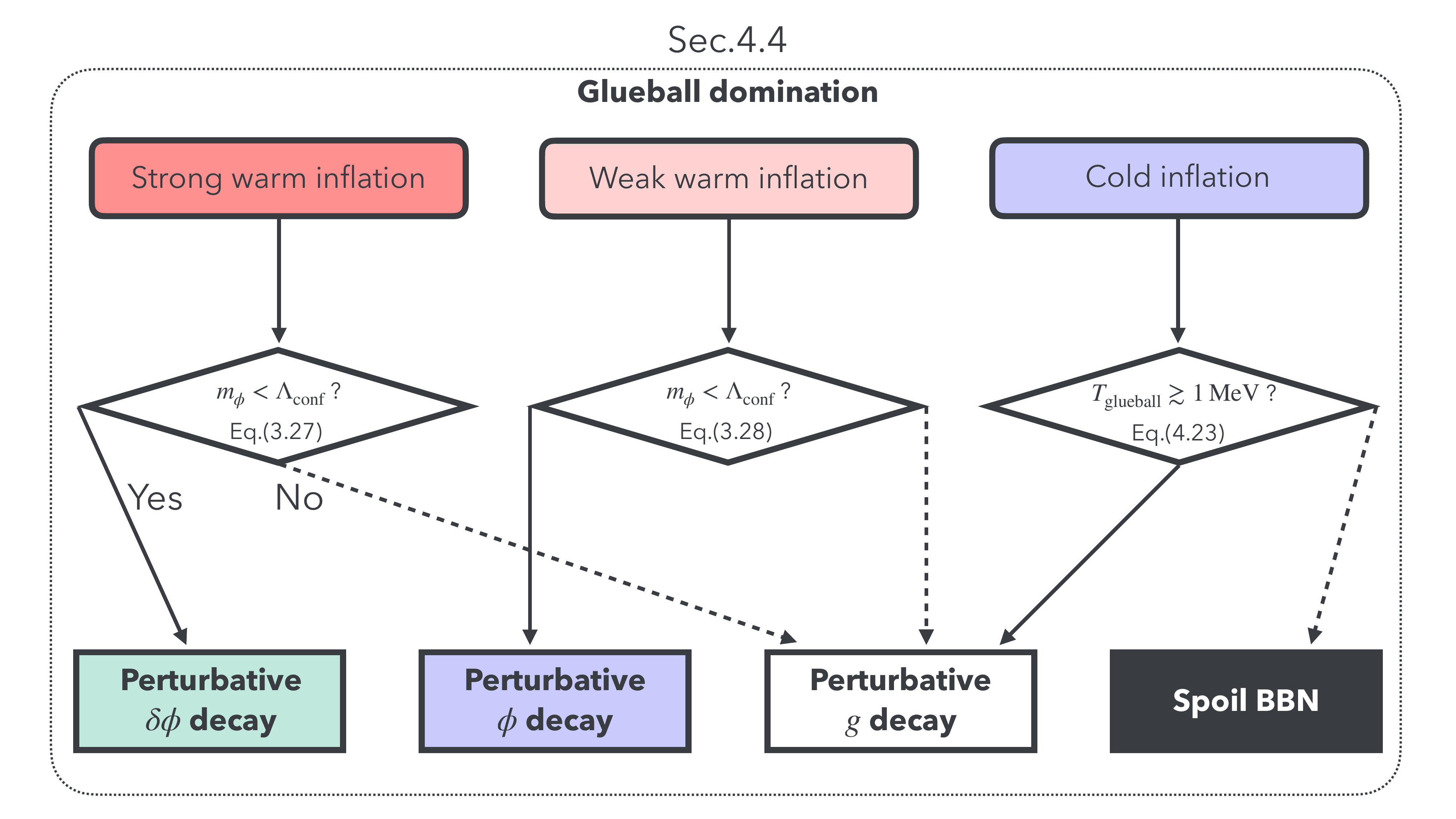}
	\caption{
    Schematic figure for the structure of this paper.
    (Top) The classification of the inflation end in Sec.~\ref{sec:endinf}.
    (Middle/Bottom) The classifications of the completion of reheating in Sec.~\ref{sec:comp_reh}.
    For the middle panel, the prompt decay of glueballs is assumed, while for the bottom panel the glueball can dominate the Universe.
    }
	\label{fig:cls}
\end{figure}

This paper is organized as follows.
In Sec.~\ref{sec:prel}, we introduce the model, the basic requirements, and the theoretical framework.
Then, we classify the phases of the inflation end in Sec.~\ref{sec:endinf}, distinguishing between warm and cold inflation.
Setting the result of Sec.~\ref{sec:endinf} as an initial condition, we classify the completion of reheating in Sec.~\ref{sec:comp_reh}.
Sec.~\ref{sec:sumdis} is devoted to summary and discussion of our findings.
See also Fig.~\ref{fig:cls}.

\section{Preliminary}
\label{sec:prel}

\subsection{Setup}
\label{sec:setup}

Throughout this paper, we consider the following simple model where the inflaton $\phi$ couples to the $\mathrm{SU}(N)$ YM theory via the CS coupling:
\begin{equation}
\label{eq:model_Lagrangian}
  \mathcal{L} = 
  \frac{1}{2} g^{\mu\nu} \partial_\mu \phi \partial_\nu \phi - V (\phi)
  - \frac{1}{4} g^{\mu \rho} g^{\nu \sigma} F^a_{\mu\nu} F_{\rho \sigma}^a
  + \frac{\phi}{8 \pi \Lambda} F^a_{\mu\nu} \tilde F^{a \mu\nu}
  + \mathcal{L}_\text{others},
\end{equation}
where the field strength of YM theory is $F^a_{\mu\nu} = \partial_\mu A^a_\nu - \partial_\nu A^a_\nu + g f_{abc} A^b_\mu A^c_\nu$ with $a = 1, \cdots, N^2 - 1$, and the dual field strength is $\tilde F^{a \mu\nu} \equiv E^{\mu\nu\rho\sigma}F^a_{\rho\sigma}/2$ with the totally antisymmetric tensor being $E^{\mu\nu\rho\sigma} = \epsilon^{\mu\nu\rho\sigma} / \sqrt{-g}$ and $\epsilon^{0123} = 1$.
Since we are interested in the background evolution of the Universe, we take the FLRW metric under a vanishing curvature, $\dd s^2 = \dd t^2 - a^2(t) \dd \bm{x}^2$, with $a (t)$ being the scale factor.
As the CS coupling to the inflaton is a higher dimensional operator, we need to make sure that the relevant scale has to be lower than the cutoff scale, which in our case implies
\begin{equation}
  \label{eq:cutoff}
  \Lambda \gtrsim \Lambda_\text{inf},
\end{equation}
with the inflation scale being $\Lambda_\text{inf} = (3 \Mpl^2 H_\text{inf}^2)^{1/4}$.
$\mathcal{L}_\text{others}$ represents the light degrees of freedom.
The role of this sector and the coupling to the YM sector will be discussed in detail in the following sections.
Throughout this paper, having in mind a typical cold high-scale inflation, we assume that the inflation end is dominated by the first slow-roll parameter [see Eq.~\eqref{eq:slowroll} for concrete definitions].
Although we are agnostic about the details of the inflaton potential, we assume that the inflaton potential is approximated by a quadratic potential
\begin{equation}
  \label{eq:potential}
  V(\phi) \simeq \frac{m_\phi^2}{2} \phi^2,
\end{equation}
after the inflation end. We take the inflaton mass to be $m_\phi = 10^{13}\,\mathrm{GeV}$ as a reference, which is a typical value after high-scale inflation.

\subsection{Confinement}
\label{sec:confinement}

The pure YM theory is asymptotically free and exhibit confinement in the IR.
The running coupling constant is evaluated by
\begin{equation}
  \label{eq:running}
  \alpha (\mu) = \frac{2 \pi}{b \ln \qty(\mu / \Lambda_\text{conf}) },\qquad
  b = \frac{11}{3} N,
\end{equation}
with $\Lambda_\text{conf}$ being the confinement scale.
Since the confinement gives rise to the additional potential via the CS coupling on top of $V(\phi)$, we need to make sure that the confinement scale is sufficiently lower than the inflation scale, \textit{i.e.,} $m_\phi^2 \gg \Lambda_\text{conf}^4 / \Lambda^2$.
Below the confinement scale, the lightest particle in the YM sector is the glueball, whose mass is estimated as $m_\text{glueball} \sim \Lambda_\text{conf}$.
The confinement phase transition is expected to be first order for $N \geqslant 3$ and second order for $N = 2$. Unless the first order phase transition is significantly supercooled, the radiation dominated Universe is expected to be realized soon after the confinement transition.
This implies that the entropy dilution factor solely from the phase transition is minor compared to the possible glueball domination afterwards, and hence we neglect the dilution by the phase transition in the following for simplicity.

Since the main interest of this paper is the process of reheating, we have to make sure that the confinement does not prevent the transition to radiation dominated Universe.
First of all, whatever happens, the non-relativistic glueballs eventually dominate the Universe unless they can also decay into radiation.
This is the reason why we introduce $\mathcal{L}_\text{others}$ that is responsible for the glueball decay.
Although the concrete decay rate depends on the details of how the YM sector couples to $\mathcal{L}_\text{others}$, we take the following higher dimensional operator as a reference, $h^2 F^{a\mu\nu} F_{\mu\nu}^a/m_X^2$ with $h$ being a light scalar field in $\mathcal{L}_\text{others}$ singlet under the $\mathrm{SU}(N)$ and $m_X$ being a high-energy scale of $m_X \gg \Lambda_\text{conf}$.\footnote{
  An operator responsible for the glueball decay can be induced as follows for instance.
  Suppose that there exists a heavy scalar $X$ charged under the $\mathrm{SU}(N)$ with its mass being $m_X > \Lambda_\text{conf}$.
  If this scalar couples to a light scalar field $h$ (singlet under the $\mathrm{SU}(N)$) in $\mathcal{L}_\text{others}$, for instance, via $X^2 h^2$, $h^2 F_{\mu\nu}^a F^{a\mu\nu} / m_X^2$ is induced below the $X$ mass.}
This operator induces the glueball decay rate into $\mathcal{L}_\text{others}$ as
\begin{equation}
\label{eq:glueball_decayrate_other}
  \Gamma_\text{glueball} \sim \frac{\Lambda_\text{conf}^5}{m_X^4}.
\end{equation}
When the inflaton mass is smaller than the glueball mass, $m_\phi < \Lambda_\text{conf}$, another decay channel opens up.
The glueball can decay into the inflaton particles via the CS coupling, whose decay rate is estimated as
\begin{equation}
  \label{eq:glueball_decayrateinto_inflaton}
  \Gamma_{\text{glueball}} \sim \frac{\Lambda_\text{conf}^5}{\Lambda^4}.
\end{equation}
This implies that the glueballs dominantly decay into the inflaton particles if $m_X \gtrsim \Lambda$.

To avoid complications, we mostly discuss the case where the glueball does not dominate the Universe, which requires $1 > (m_\text{glueball}/T_\text{glueball} ) (\rho_\text{glueball} / \rho_\text{rad})_{T \sim m_\text{glueball}}$ with $T_\text{glueball}$ being the temperature of the glueball decay.\footnote{
  Strictly speaking, the glueball number density freezes out when the number violating processes such as $ggg \rightarrow gg$ become inefficient, which occurs at $T \sim m_\text{glueball}/20$, a bit later than $T \sim m_\text{glueball}$.
  We do not take this additional suppression via the freeze-out into account for simplicity, and assume that the glueball number density decays in proportional to $a^{-3}$ after $T \lesssim m_\text{glueball}$.
}
This condition implies the following lower bound on the glueball decay rate, or equivalently the upper bound on $m_X$ or $\Lambda$:
\begin{equation}
  \label{eq:prompt_glueball_decay}
  \Gamma_\text{glueball} \gtrsim g_\ast^{-2}\, \qty( \frac{\pi^2 g_\ast}{90} )^{1/2}\, \frac{\Lambda_\text{conf}^2}{\Mpl} ~\longrightarrow~
  g_\ast^{1/2}\, \qty( \frac{\pi^2 g_\ast}{90} )^{-1/8} \qty( \Mpl \Lambda_\text{conf}^3 )^{1/4} \gtrsim
  \begin{cases}
  m_X & \text{for}\quad m_X \lesssim \Lambda, \\
  \Lambda  & \text{for}\quad m_X \gtrsim \Lambda. \\
  \end{cases}
\end{equation}
To avoid unnecessary complications caused by the presence of $X$, we take
\begin{equation}
  \label{eq:mX_max}
  m_X = \bar{m}_{X} \equiv g_\ast^{1/2}\, \qty( \frac{\pi^2 g_\ast}{90} )^{-1/8} \qty( \Mpl \Lambda_\text{conf}^3 )^{1/4} ,
\end{equation}
for the case of no glueball domination.
We separately discuss the opposite case where the glueball once dominates the Universe in Sec.~\ref{sec:gluedom}.

In addition, the inflaton decay rate also depends on the competition between $m_\phi$ and $\Lambda_\text{conf}$.
For $m_\phi > \Lambda_\text{conf}$, the inflaton can directly decay into the YM sector, whose decay rate is solely determined by the CS coupling [see Eq.~\eqref{eq:vac_decay}].
For $m_\phi < \Lambda_\text{conf}$, while the inflaton cannot decay into the YM sector directly, it can decay into $\mathcal{L}_\text{others}$ via a higher-dimensional operator obtained from integrating out the heavy glueball.
The decay rate of inflaton in this case, however, depends on a particular modeling of $\mathcal{L}_\text{others}$.
As a reference, we again take $h^2 F^{a\mu\nu} F_{\mu\nu}^a/m_X^2$, which yields
\begin{equation}
  \label{eq:prompt_glueball_decay_inflaton_decay}
  \Gamma_\phi \sim \frac{\Lambda_\text{conf}^8}{\Lambda^2 m_X^4 m_\phi},
\end{equation}
after integrating out the heavy glueballs.

For later convenience, we rewrite Eq.~\eqref{eq:running} as follows:
\begin{equation}
  \label{eq:confscale}
  \Lambda_\text{conf} = \mu \exp \qty( - \frac{2 \pi}{b \alpha (\mu)} ).
\end{equation}
For a given $\alpha (\mu)$ specified at a UV scale $\mu$, we can estimate the confinement scale $\Lambda_\text{conf}$ by using this equation. This expression is only true for the pure YM theory, and the actual confinement scale can depend on $m_X$ if $m_X < \Lambda_\text{inf}$.
The amount of modification on the confinement scale can be estimated as
\begin{equation}
  \frac{\Lambda_{\text{conf},X}}{\Lambda_\text{conf}}
  \simeq
  \qty( \frac{\Lambda_\text{inf}}{m_X} )^{\frac{b_X}{b}},
\end{equation}
where the beta function coefficient of $X$ is $b_X = n_X T_{R_X} / 3$ for scalar $X$ with $n_X$ being the number of $X$ under the representation $R_X$ in the $\mathrm{SU}(N)$, and $T_{R_X}$ being the Dynkin index of $R_X$.
For instance, for a single scalar field under the fundamental representation, \textit{i.e.}, $n_X = 1$ and $T_{R_X} = 1/2$, the exponent is tiny, $b_X / b = 1/(22 N)$.
Therefore, unless we introduce $X$ with large degrees of freedom or $m_X$ is significantly lower than $\Lambda_\text{inf}$, the effect of $X$ on modifying the confinement scale by $X$ can be safely neglected, which is the case in the following discussion.

\subsection{Key quantities}

In general, we need to solve the field equations of the inflaton and gauge fields simultaneously.
However, in the most cases of our interest, as we will see in the following (see Sec.~\ref{sec:endinf} and Sec.~\ref{sec:comp_reh}), one may safely approximate that the YM sector is thermalized, which dramatically simplifies the analysis. 
The thermalization rate of YM plasma is estimated as follows
\begin{equation}
  \label{eq:thermalization}
  \Gamma_\text{th} \sim c N^2 \alpha^2 T,
\end{equation}
with a numerical factor being $c \sim 10$ according to ~\cite{Laine:2021ego}.
If the thermalization of YM plasma within the relevant time scale is guaranteed, one may trace out (or integrate out) the gauge fields, yielding the dynamical equation only consists of the homogeneous inflaton condensate $\phi(t)$ and its fluctuations $\delta \phi$.
The relevant equations of inflaton in our case are the followings:
\begin{align}
  &0 = \ddot \phi (t) + \qty( 3 H + \Gamma_\phi ) \dot \phi(t) + V', \label{eq:eom_homphi} \\
  &\qty(\partial_t - H k \partial_{k} ) f_{\delta \phi} (t, k)
  = \Gamma_{\delta \phi} (k) \qty[ f_\text{B} (k) - f_{\delta \phi} (t,k) ].
\end{align}
Here the Hubble parameter is $H \equiv \dot a / a$, the inflaton phase-space density is denoted by $f_{\delta \phi}$, and the Bose--Einstein distribution function is given by $f_\text{B} (k) \equiv 1 / (e^{\omega_{k}/T} - 1)$ with $\omega_k \equiv (k^2 + m_\phi^2)^{1/2}$.
The friction term $\Gamma_\phi$ for the inflaton condensate and the production rate of inflaton particles $\Gamma_{\delta \phi}$ stem from the same correlator but in different regions of momentum space:
\begin{align}
  \Gamma_\phi &= \frac{1}{64 \pi^2 \Lambda^2} \frac{\Pi_\phi^\rho (m_\phi,0)}{2 m_\phi}, \\
  \Gamma_{\delta \phi} (k) &= \frac{1}{64 \pi^2 \Lambda^2} \frac{\Pi_\phi^\rho (\omega_k,k)}{2 \omega_k},
\end{align}
where the spectral function for inflaton is defined by
\begin{equation}
  \Pi^\rho_\phi (t , {\bm x}) \equiv \left< \qty[ \{ F_{\mu\nu}^a \tilde F^{a \mu \nu} \} (t ,{\bm x}), \{ F_{\mu\nu}^a \tilde F^{a \mu \nu} \} (0 ,{\bm 0}) ]\right>, \quad
  \Pi^\rho_\phi (\omega, k) \equiv \int \dd t \dd {\bm x} \, e^{i \omega t - i {\bm k} \cdot {\bm x}}\, \Pi^\rho_\phi (t , {\bm x}).
\end{equation}
The expectation value is taken over the thermal state for the YM sector, which implies the translational invariance and isotropy of the correlator.

Let us first consider when the temperature of the Universe is sufficiently low, $m_\phi \gg T$.
In this case, the finite-temperature corrections to the friction and the production rate can be safely neglected.
Also, the inflaton particles would be mostly dominated by the non-relativistic component.
Hence the friction for the condensate and the production rate are given by the inflaton decay rate evaluated at vacuum, which is nothing but the well-known axion decay rate into gauge bosons:
\begin{equation}
  \label{eq:vac_decay}
  \gamma_k \Gamma_{\delta\phi} (k) \simeq \Gamma_\phi 
  \simeq
  \frac{d_\text{Ad}}{256 \pi^3} \frac{m_\phi^3}{\Lambda^2} \qquad \text{for} \quad m_\phi > \Lambda_\text{conf}.
\end{equation}
Here the boost factor is $\gamma_k \equiv \omega_k / m_\phi$ and the dimension of the adjoint representation is $d_\text{Ad} = N^2 - 1$ for $\mathrm{SU}(N)$ gauge theory.
For $m_\phi < \Lambda_\text{conf}$, it is instead given by Eq.~\eqref{eq:prompt_glueball_decay_inflaton_decay}.

When the temperature is much higher than the inflaton mass and the confinement scale, $T \gg m_\phi$ and $T \gg \Lambda_\text{conf}$, the finite-temperature corrections significantly alter the friction and production rate.
The production of inflaton particles by the thermal YM plasma is dominated by $k \sim T$ as the typical momentum of gauge bosons is $T$.
For $k \gtrsim T$, one may compute the rate perturbatively~\cite{Masso:2002np,Graf:2010tv,Salvio:2013iaa,Bouzoud:2024bom}
\begin{equation}
  \label{eq:inflaton_production}
  \overline{\Gamma}_{\delta\phi} \equiv \frac{\int_{\bm k} f_\text{B}(k)\Gamma_{\delta\phi}(k)}{\int_{\bm k} f_\text{B}(k)}
  =
  \frac{d_\text{Ad}}{16 \pi^2} \frac{N}{3}  \frac{\alpha T^3}{\Lambda^2} f(T),
\end{equation}
with $f (T)$ being an order one function.
On the other hand, for $m_\phi, k  \ll \alpha^2 T$, the calculation based on the quasi-particle of gluon is broken down, and we have to deal with hydrodynamic excitations.
In our case, the relevant rate is imprinted in the CS diffusion via the sphaleron at high temperature.
For this reason, the thermal friction of inflaton condensate is given
by~\cite{McLerran:1990de,Co:2019wyp,Berghaus:2019whh,Domcke:2020kcp,Laine:2021ego,Drewes:2023khq}\footnote{
  This expression is only valid in the absence of any emergent (approximate) conservation law in the thermal plasma involving the axion shift charge. Regarding the axion-inflaton potential as an external force driving the charge production, the friction term is interpreted as a resistance of the thermal plasma trying to washout the non-zero charge. Hence, if the axion shift charge is a part of a conserved charge in the thermal plasma, the effective friction term vanishes as there is no resistance.
  If the axion shift charge is a part of an approximate conserved charge, the effective friction term can be suppressed by the bottleneck process of washout.
  See \textit{e.g.,} \cite{McLerran:1990de, Co:2019wyp,Berghaus:2019whh,Domcke:2020kcp, Drewes:2023khq} for more details. In our setup, no such conserved combination arises, so we can safely use the friction term.
}
\begin{equation}
  \label{eq:thermal_fric}
  \Gamma_\phi \sim \frac{d_\text{Ad} N^3}{4} \frac{\alpha^3 T^3}{\Lambda^2} \qquad \text{for} \quad m_\phi \ll \Gamma_\text{th}.
\end{equation}

\section{End of Inflation}
\label{sec:endinf}

The CS coupling can efficiently produce gauge fields even during inflation. Hence, we have to consider its effects before the end of inflation, even though our main focus is on the completion of reheating.
We first discuss the production of gauge fields during inflation when the thermal YM plasma is present from the beginning, and then extend the discussion to the case where there are no initial gauge fields.
The main purpose of this section is to perform a comprehensive analysis of the amount of gauge fields right after inflation.
The result of this section is summarized in Figs.~\ref{fig:cls} and \ref{fig:infend}.

\subsection{Warm axion inflation}
Suppose that the thermal YM plasma somehow exists during inflation, which at least requires
\begin{equation}
  \label{eq:thermcond}
  \Gamma_\text{th} \gtrsim H,
\end{equation}
for self-consistency.
As we see shortly, once such thermal YM plasma is somehow generated, the system is driven to an attractor solution known as the warm inflation.

\vskip.3em
\noindent
\textbf{Basin of attraction.---}
As long as the thermalization is faster than the system evolution, the gauge fields generated within a Hubble time are quickly thermalized.
Since the previously generated gauge fields are diluted by the cosmic expansion, the newly produced ones always dominate.
This balance determines the evolution of the radiation energy density as follows
\begin{equation}
  \label{eq:rho_rad_w}
  \dot \rho_\text{rad} + 4 H \rho_\text{rad} = \Gamma_\phi \dot \phi^2 ~\longrightarrow ~
  \rho_\text{rad} \simeq \frac{\Gamma_\phi \dot \phi^2}{4 H},
\end{equation}
where the energy density of radiation is given by
\begin{equation}
  \rho_\text{rad} = \frac{\pi^2 g_\ast}{30} T^4, \qquad
  g_\ast = g_\text{YM} + g_\text{others} + g_\phi,
\end{equation}
with $g_\text{YM}$, $g_\text{others}$, and $g_\phi$ representing the relativistic degrees of freedom of the YM sector, $\mathcal{L}_\text{others}$, and inflaton particle, respectively.
Even if the system initially has temperature smaller than this equilibrium solution, the temperature quickly increases within the cosmic time as $T \sim t \alpha^3 \dot \phi^2 / \Lambda^2$ as one can show from $\dot \rho_\text{rad} \sim \Gamma_\phi \dot \phi^2$.
Hence, within $t \lesssim H^{-1}$, it is clear that the equilibrium solution \eqref{eq:rho_rad_w} is realized because the initial $\dot \phi$ is larger for a smaller temperature.

In the presence of the thermal YM plasma, the inflaton condensate acquires the friction, which modifies the slow-roll parameters
\begin{equation}
  \label{eq:slowroll}
  \epsilon_T = \frac{1}{1 + Q} \epsilon_V, \qquad
  \eta_T = \frac{1}{1 + Q} \eta_V, \qquad
  \epsilon_V = \frac{\Mpl^2}{2} \qty( \frac{V'}{V} )^2, \qquad
  \eta_V = \Mpl^2 \frac{V''}{V}, \qquad Q \equiv \frac{\Gamma_\phi}{3 H},
\end{equation}
where the slow-roll condition is $\epsilon_T, |\eta_T| < 1 $.
Under the modified slow-roll condition, one finds that the equation of motion given in Eq.~\eqref{eq:eom_homphi} is reduced to
\begin{equation}
  \label{eq:slowroll_w}
  \dot \phi \simeq - \frac{1}{1 + Q} \frac{V'}{3 H}.
\end{equation}
Plugging Eq.~\eqref{eq:slowroll_w} into Eq.~\eqref{eq:rho_rad_w}, we find the attractor solution for the energy density of radiation
\begin{equation}
  \label{eq:rho_rad_infend}
  \rho_\text{rad} \simeq \frac{\epsilon_T}{2} \frac{Q}{1 + Q} \, V
  ~\xrightarrow{\text{end}}~
  \frac{1}{2} \frac{Q_\text{end}}{1 + Q_\text{end}} \, V_\text{end},
\end{equation}
where the arrow $\xrightarrow{\text{end}}$ and the subscript ``end'' indicate the inflation end.

\vskip.3em
\noindent
\textbf{Self consistency.---}
When the condition of the weak warm inflation, \textit{i.e.,} $Q < 1$, is fulfilled till the end of inflation, the temperature of radiation can be estimated from Eq.~\eqref{eq:rho_rad_infend} as
\begin{equation}
  \label{eq:Tweak}
  T \sim \epsilon_V \frac{d_\text{Ad} N^3}{8 \sqrt{3}} \qty(\frac{\pi^2 g_\ast}{30})^{-1} \alpha^3 \frac{\Mpl V^{1/2}}{\Lambda^2} 
  ~\xrightarrow{\text{end}}~
  \frac{d_\text{Ad} N^3}{8 \sqrt{3}} \qty(\frac{\pi^2 g_\ast}{30})^{-1} \alpha^3 \frac{\Mpl V^{1/2}_\text{end}}{\Lambda^2}.
\end{equation}
For the self consistency, we require the thermalization condition \eqref{eq:thermalization} and \eqref{eq:thermcond} is satisfied \textit{at least} around the end of inflation, yielding an upper bound on the cutoff scale:
\begin{equation}
  \label{eq:warm_thermalization}
  \Lambda \lesssim \Lambda_\text{w-th} \equiv
  \frac{d_\text{Ad}^{1/2} N^{5/2}}{2 \sqrt{2}} \qty(\frac{\pi^2 g_\ast}{30})^{-1/2} c^{1/2} \alpha^{5/2} \Mpl.
\end{equation}
The inequality is saturated for $\Gamma_\text{th} \simeq H$ at the end of inflation. 
If the cutoff scale exceeds this bound, the YM plasma is no longer assumed to be thermalized. Note again that this bound is only a necessary condition for the thermalization of the YM plasma. In reality, the thermalization condition is even more stringent, as it must be satisfied throughout inflation rather than solely at its end.

On the other hand, if the system enters the strong warm inflation, \textit{i.e.,} $Q > 1$, before the end of inflation, the thermalization condition is automatically fulfilled as long as $\alpha \ll 1$.
From Eq.~\eqref{eq:rho_rad_infend}, the temperature of radiation reads
\begin{equation}
  \label{eq:Tstr}
  T \sim \epsilon_T^{1/4} \qty( \frac{\pi^2 g_\ast}{15} )^{-1/4} V^{1/4}
  ~\xrightarrow{\text{end}}~
  \qty( \frac{\pi^2 g_\ast}{15} )^{-1/4} V^{1/4}_\text{end}.
\end{equation}
For self consistency, the strong warm inflation condition, $Q > 1$, must be fulfilled for Eq.~\eqref{eq:Tstr} at least at the end of inflation, which provides a boundary of the weak and strong warm inflation as follows:
\begin{equation}
  \label{eq:wbdry}
  \Lambda \lesssim
  \Lambda_\text{w-bdry} \equiv
  \frac{d_\text{Ad}^{1/2} N^{3/2}}{2 \cdot 3^{1/4}} \qty( \frac{\pi^2 g_\ast}{15} )^{-3/8} \alpha^{3/2} \Mpl^{1/2} V^{1/8}_\text{end},
\end{equation}
which is fulfilled in the strong warm inflation. The weak warm inflation is realized for $\Lambda \gtrsim \Lambda_\text{w-bdry}$.

In the strong warm inflation, the potential energy of inflaton asymptotes to a certain value at the end of inflation depending on the cutoff scale of the CS coupling, $\Lambda$.
The slow-roll condition implies $\epsilon_V \sim Q \gg 1$ at the end of inflation. If the inflaton potential is approximated by a quadratic one $\phi^2$ at the inflation end, the potential slow-roll condition can be expressed as $\epsilon_V \simeq 2 \Mpl^2 / \phi_\text{end}^2$.
By using Eqs.~\eqref{eq:thermal_fric} and \eqref{eq:Tstr}, one may rewrite $Q$ and equate it with $\epsilon_V$, which leads to
\begin{equation}
  V_\text{end}^{1/4} \sim 3^{1/10} 2^{2/5} d_\text{Ad}^{-1/5} N^{-3/5} \qty( \frac{\pi^2 g_\ast}{15} )^{3/20} \alpha^{-3/5} \qty( m_\phi^2 \Mpl )^{1/5} \Lambda^{2/5}.
\end{equation}
This implies that the temperature of the strong warm inflation does not explicitly depend on $V$.
The temperature at the end of the strong warm inflation can be estimated as
\begin{equation}
  \label{eq:Tend_str}
  T_\text{end} \sim 3^{1/10} 2^{2/5} d_\text{Ad}^{-1/5} N^{-3/5} \qty( \frac{\pi^2 g_\ast}{15} )^{-1/10} \alpha^{-3/5}\, \qty( m_\phi^2 \Mpl )^{1/5} \Lambda^{2/5}.
\end{equation}
Also, the condition given in Eq.~\eqref{eq:cutoff} puts a lower bound on the cutoff scale
\begin{equation}
  \Lambda \gtrsim 
  2^{2/3} 3^{1/6} d_\text{Ad}^{-1/3} N^{-1} \qty( \frac{\pi^2 g_\ast}{15} )^{1/4} \alpha^{-1} \qty( m_\phi^2 \Mpl )^{1/3}.
\end{equation}

\subsection{Warming up cold axion inflation}

In the beginning of the previous subsection, we have shown that the warm inflation is an attractor for an inflation model given in Eq~\eqref{eq:model_Lagrangian} as long as there exists initial thermal plasma fulfilling Eq.~\eqref{eq:thermcond}.
This consideration basically implies that the thermalization of YM sector is expected if the energy density exceeds the following inequality:
\begin{equation}
  \label{eq:thermalization_warmingup}
  \rho_\text{rad} \gtrsim \qty( \frac{\pi^2 g_\ast}{30} ) \qty( \frac{H}{c N^2 \alpha^2} )^4.
\end{equation}
In this section, we consider the inflation starting from vacuum of the YM plasma, namely the cold inflation.\footnote{
  The $X$ matter, which is responsible for the glueball decay, may participate in the thermal plasma in a certain parameter of our interest.
  However, unless the degrees of freedom of $X$ are significantly large, the effect of $X$ on the thermalization of YM sector is minor.
  Hence we neglect the presence of $X$ in the following discussion.
}
We discuss the condition where a sufficient amount of radiation satisfying Eq.~\eqref{eq:thermalization_warmingup} is generated from vacuum, following Ref.~\cite{DeRocco:2021rzv}.

\vskip.3em
\noindent
\textbf{Sufficient production.---}
Suppose that there are no initial gauge fields during inflation.
Before the gauge field amplitude becomes so large that the non-linear interactions due to the non-Abelian nature cannot be neglected, one may approximate the equation of motion at a linear order in the gauge field~\cite{Turner:1987bw,Garretson:1992vt,Anber:2006xt,Fujita:2015iga,Adshead:2016iae}
\begin{equation}
  \ddot A_{\pm}^a + k \qty(k \pm \frac{\dot \phi}{2\pi \Lambda} ) A_{\pm}^a = 0,
\end{equation}
with the subscripts $\pm$ representing two circular polarizations.
Here we have neglected the cosmic expansion, which is sufficient for our purpose.
Depending on the sign of the velocity of $\phi$, one may readily see that one particular mode exhibits the tachyonic instability, \textit{i.e.,} $A_\pm$ grows exponentially for $\dot \phi \lessgtr 0$ respectively, whose exponent is controlled by $k_\text{inst} t$ with $k_\text{inst} \equiv |\dot \phi| / (2 \pi \Lambda)$.
In order to generate gauge fields fulfilling Eq.~\eqref{eq:thermalization_warmingup}, the production must be efficient within the one Hubble time, which implies
\begin{equation}
  \label{eq:sufficient_prod}
  k_\text{inst} \gg H ~ \longrightarrow ~
  \Lambda \ll \mathcal{O} (0.1) \epsilon_V^{1/2} \Mpl.
\end{equation}
Once this condition is met, the self-interactions of the gauge fields cannot be neglected, which terminates the exponential growth.
The amplitude of the gauge field where the self-interactions come into play can be estimated as
\begin{equation}
  \label{eq:terminalA}
  A_\text{NL} \sim \frac{k_\text{inst}}{2 g N},
\end{equation}
by comparing $\partial^2 A^a$ with $g f_{abc} A^b \partial A^c$.
\footnote{
  Strictly speaking, to reach this field value within one Hubble time, we expect a logarithmic correction to Eq.~\eqref{eq:sufficient_prod}. Within the range of the gauge coupling of our interest, this correction is irrelevant for our order of magnitude estimation.
  See Appendix of Ref.~\cite{DeRocco:2021rzv} for details.
}
When the exponential growth stops by the self-interactions, the terminal energy density of the gauge fields can be estimated as
\begin{equation}
  \label{eq:terminalrho}
  \rho_\text{rad}^\text{NL} \sim d_\text{Ad} \frac{k_\text{inst}^2}{4} A_\text{NL}^2
  \sim \frac{d_\text{Ad}}{N^2} \frac{H^4}{64 \pi \alpha} \times \qty( \frac{\sqrt{\epsilon_V}}{\sqrt{2} \pi} \frac{\Mpl}{\Lambda} )^4.
\end{equation}
Here we have used the terminal amplitude of the gauge field given in Eq.~\eqref{eq:terminalA}.
If this energy density exceeds the condition given in Eq.~\eqref{eq:thermalization_warmingup}, the YM plasma generated via the tachyonic instability is thermalized
\begin{equation}
  \label{eq:nl_thermalization}
  \Lambda \ll \frac{15^{1/4}c}{2^{7/4} \pi^{7/4}} d_\text{Ad}^{1/4} g_\ast^{-1/4} N^{3/2} \alpha^{7/4} \epsilon_V^{1/2} \Mpl
  \sim N^{3/2} \alpha^{7/4} \epsilon_V^{1/2} \Mpl.
\end{equation}

We require the conditions Eqs.~\eqref{eq:sufficient_prod} and \eqref{eq:nl_thermalization} for the warm-inflation regime:
\begin{equation}
  \label{eq:warmedup}
  \Lambda < \Lambda_\text{warmed-up}, \qquad
  \Lambda_\text{warmed-up} \equiv \epsilon_V^{1/2} \Mpl \min \qty[\mathcal{O} (0.1),   \frac{15^{1/4}c}{2^{7/4} \pi^{7/4}} d_\text{Ad}^{1/4} g_\ast^{-1/4} \, N^{3/2} \alpha^{7/4}].
\end{equation}
This subtlety does not alter the process of reheating as we will see shortly.

\vskip.3em
\noindent
\textbf{Backreaction to inflaton.---}
If the backreaction significantly alters the inflaton motion before the thermalization of gauge fields comes into play, a strong non-equilibrium behavior observed in $\mathrm{U}(1)$ gauge theories would be expected~\cite{Cheng:2015oqa,Domcke:2020zez,Caravano:2022epk,Gorbar:2021rlt,Caravano:2022epk,Figueroa:2023oxc,Domcke:2023tnn}.
In this case, we cannot rely on the thermalization of the YM plasma, and hence 
studies using numerical lattice simulations become unavoidable.
Here we derive the condition to avoid the significant backreaction to the inflation motion. To sustain the slow-roll motion of the inflaton, the energy density of the gauge field produced by tachyonic instability should not exceed the kinetic energy density of the inflaton condensate.
We need to ensure that the gauge field can reach the terminal energy density of Eq.~\eqref{eq:terminalrho} without violating this requirement, which implies
\begin{equation}
\rho_\text{rad}^\text{NL}\ll \rho_\phi ^\text{kin}
 ~ \longrightarrow ~\mathcal{O} (10^{-5})\frac{\epsilon_V}{\alpha}V\ll \Lambda^4
\end{equation}
where $\rho_\phi ^\text{kin}$ represents as kinetic energy density of the inflaton.
This inequality is satisfied in the parameters of our interest, and hence we can safely neglect the backreaction to inflaton.

\begin{figure}[t]
	\centering
  \hfill
  \includegraphics[height=0.27\linewidth]{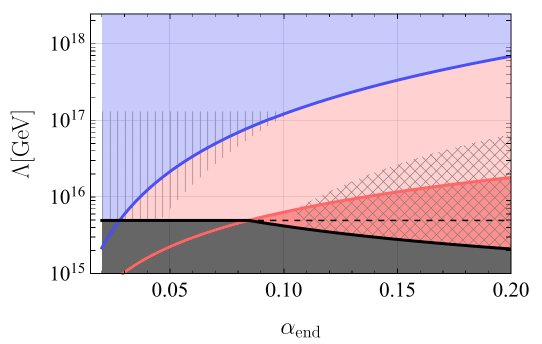}
  \hfill
  \includegraphics[height=0.27\linewidth]{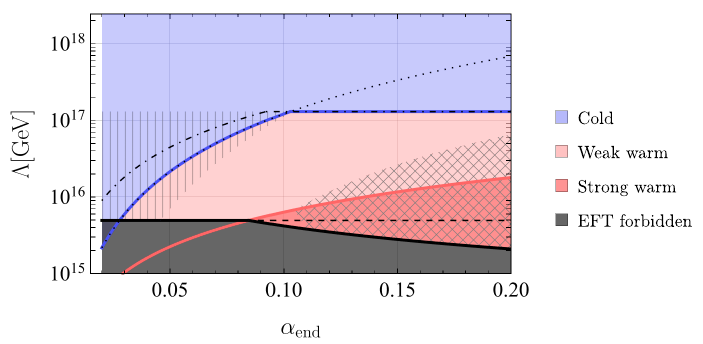}
  \hfill
	\caption{
    The phases of axion inflation at the inflation end on the $(\alpha_\text{end},\Lambda)$ plane for $m_\phi = 10^{13}\,\mathrm{GeV}$, $V^{1/4}_\text{end} = 5 \times 10^{15}\,\mathrm{GeV}$ (for weak warm/cold inflation) and $N=3$; in the presence/absence initial thermal YM plasma (Left/Right) respectively.
    The gauge coupling, $\alpha_\text{end}$, is evaluated by the temperature at the end of inflation, $T_\text{end}$ (warm inflation), or the Hubble parameter at the end of inflation $H_\text{end}$ (cold inflation).
    Each inflation phase is filled in blue (cold), pink (weak warm), and red (strong warm).
    The black shaded region corresponds to the parameters outside the validity of EFT, $\Lambda_\text{inf} > \Lambda$, at the end of inflation.
    If we require $\Lambda_\text{inf} < \Lambda$ throughout the whole inflation, this bound could be as strong as the black dashed line.
    In the right panel, we have shown the dotted (dot-dashed) line, $\Lambda = \Lambda_\text{w-th}$ ($\Lambda_\text{warmed-up}$), to clarify that the weak warm inflation requires both conditions.
    The reheating of each phase will be discussed in the next section; strong warm (red) in Sec.~\ref{sec:strw_reh}, and weak warm (pink) in Sec.~\ref{sec:ww_reh}, and cold (blue) in Sec.~\ref{sec:c_reh}, respectively.
    The hatched region, fulfilling $k_\text{inst} > m_\text{M},m_\phi$, requires discussion on preheating as an efficient tachyonic instability is present (see the second subsubsections of Secs.~\ref{sec:ww_reh} and \ref{sec:c_reh}).
    The cross-hatched region involves an additional model dependence of reheating because the confinement scale exceeds the inflaton mass, \textit{i.e.,} $\Lambda_\text{conf} > m_\phi$.}
	\label{fig:infend}
\end{figure}

\vskip.3em
\noindent
\textbf{Warm inflation attractor.---}
Up to here, we have discussed the conditions under which the gauge fields generated from vacuum are expected to be thermalized within the Hubble time.
However, this does not immediately mean that the system falls into the attractor solution of the warm inflation given in Eq.~\eqref{eq:rho_rad_infend}.
The tachyonic instability is terminated if the instability scale is smaller than the magnetic screening mass, $k_\text{inst} \lesssim m_\text{M}$~\cite{Mukaida:2012qn,Mukaida:2012bz,Enqvist:2012tc,Lerner:2015uca} with $m_\text{M} \simeq g^2 N T$~\cite{Teper:1998te,Laine:2009dh}.
If this condition is fulfilled for the temperature determined by the warm inflation attractor, one may safely neglect the tachyonic instability, whose condition is
\begin{equation}
  \Lambda \lesssim
  \frac{\pi^2}{\sqrt{2}} d_\text{Ad} N^4 \qty( \frac{\pi^2 g_\ast}{30} )^{-1} \alpha^4 \Mpl.
\end{equation}
Here we have used the temperature for the weak warm inflation because the tachyonic instability is not efficient in the regime of strong warm inflation, and this equality is satisfied if $\epsilon_T = 1$.
This upper bound on $\Lambda$ becomes stronger than Eq.~\eqref{eq:warmedup} for a smaller $\alpha$ (see Fig.~\ref{fig:infend}).

For $k_\text{inst} \gtrsim m_\text{M}$, the tachyonic instability would be still present even if $\Gamma_\text{th}> H$ because the instability scale exceeds the thermalization rate, $k_\text{inst} > \Gamma_\text{th}$, implying a possible exponential growth within $1/\Gamma_\text{th}$.
As long as the condition \eqref{eq:nl_thermalization} is satisfied, the gauge fields are thermalized quickly within $1/\Gamma_\text{th}$ when the amplitude grows as large as $A_\text{NL}$, where the non-linear effects become relevant.
This process continuously generates the gauge fields with the amount of Eq.~\eqref{eq:terminalrho} within $1/\Gamma_\text{th}$, and eventually, the production is balanced by the dilution of the cosmic expansion, leading to
\begin{equation}
  \rho_\text{rad} \sim \rho_\text{rad}^\text{NL} \times \frac{\Gamma_\text{th}}{H}
  \sim \frac{c}{2^{11} 3^2 \pi^5} d_\text{Ad}^2 N^3 \, \qty( \frac{\pi^2 g_\ast}{30} )^{-1} \alpha^4 \, \qty( \frac{\Mpl}{\Lambda} )^6 \frac{V}{\Mpl^4} \times V,
\end{equation}
at the end of inflation, \textit{i.e.,} $\epsilon_V = 1$.
If this energy density is smaller that that generated by the warm inflation attractor, one can safely neglect this contribution.
This situation is realized for
\begin{equation}
  \Lambda \lesssim 10^{18} \, \mathrm{GeV} \times \Big( \frac{d_\text{Ad}}{8} \Big)^{1/5} \qty( \frac{N}{3} )^{9/5} \qty( \frac{\alpha}{0.1} )^{8/5},
\end{equation}
which is automatically satisfied in the parameters of our interest (See Fig.~\ref{fig:infend}).
Therefore, we conclude that the warm inflation attractor is robust even if the tachyonic instability is not completely turned off.
The hatched-pink region in Fig.~\ref{fig:infend} corresponds to the parameters where the tachyonic instability is present but the gauge fields are still governed by the warm inflation attractor.

\subsection{Confinement after inflation}

Before closing this section, we briefly comment on the confinement scale.
So far, we have discussed the warm inflation attractor, where the gauge coupling should be evaluated at the temperature of the YM plasma, given in Eqs.~\eqref{eq:Tweak} and \eqref{eq:Tend_str}.
For a given $\Lambda$ and $\alpha_\text{end} = \alpha (T_\text{end})$ evaluated at the inflation end, one may estimate the confinement scale by using Eq.~\eqref{eq:confscale}:
\begin{equation}
  \Lambda_\text{conf} \sim
  \exp \qty( - \frac{2 \pi}{b \alpha_\text{end}} ) 
  \begin{cases}
    \label{eq:conf_end}
    3^{1/10} 2^{2/5} d_\text{Ad}^{-1/5} N^{-3/5} \qty( \frac{\pi^2 g_\ast}{15} )^{-1/10} \alpha^{-3/5}\, \qty( m_\phi^2 \Mpl )^{1/5} \Lambda^{2/5}
    &\text{for strong warm}, \\[.5em]
    \frac{d_\text{Ad} N^3}{8 \sqrt{3}} \qty(\frac{\pi^2 g_\ast}{30})^{-1} \alpha^3_\text{end} \frac{\Mpl V^{1/2}_\text{end}}{\Lambda^2}
    &\text{for weak warm}.
  \end{cases}
\end{equation}
For cold inflation, the gauge fields are not yet thermalized at the end of inflation.
In this case, we estimate the confinement scale by fixing the gauge coupling at the horizon scale, \textit{i.e.,} $\alpha_\text{end} = \alpha (H_\text{end})$, which gives
\begin{equation}
  \label{eq:conf_end_c}
  \Lambda_\text{conf} \sim
  \exp \qty( - \frac{2 \pi}{b \alpha_\text{end}} )\, H_\text{end} \quad\text{for cold}.
\end{equation}

As explained at the end of Sec.~\ref{sec:setup}, the decay rate of inflaton is no longer described by Eq.~\eqref{eq:vac_decay} if $m_\phi < \Lambda_\text{conf}$, where the dynamics depends on a specific modeling of $\mathcal{L}_\text{others}$.
This happens at 
\begin{equation}
  \label{eq:conf_str}
    \Lambda \gtrsim 
    3.7 \times 10^{11} \,\mathrm{GeV}\, \Big( \frac{d_\text{Ad}}{8} \Big)^{1/2} \Big( \frac{N}{3} \Big)^{3/2} \Big( \frac{g_\ast}{30} \Big)^{1/4} \Big( \frac{m_\phi}{10^{13}\,\mathrm{GeV}} \Big)^{3/2} \alpha_\text{end}^{3/2}\, e^{ \frac{15 \pi}{11 N \alpha_\text{end}} },
\end{equation}
for strong warm, and
\begin{equation}
  \label{eq:conf_w}
    \Lambda \lesssim 3.5 \times 10^{18} \,\mathrm{GeV} \, \Big( \frac{d_\text{Ad}}{8} \Big)^{1/2} \Big( \frac{N}{3} \Big)^{3/2} \Big( \frac{g_\ast}{30} \Big)^{-1/2} \Big( \frac{m_\phi}{10^{13}\,\mathrm{GeV}} \Big)^{-1/2} \Big( \frac{V^{1/4}_\text{end}}{5\times 10^{15}\,\mathrm{GeV}} \Big)\, \alpha_\text{end}^{3/2} e^{- \frac{3 \pi}{11 N \alpha_\text{end}} },
\end{equation}
for weak warm.
See the cross-hatched regions in Fig.~\ref{fig:infend}.

\section{Completion of Reheating}
\label{sec:comp_reh}

In the previous section, we have seen that there exists a non-negligible amount of gauge fields already at the end of inflation in various cases.
In this section, we discuss the reheating after inflation, particularly focusing on the completion of reheating.
As we see shortly, the existence of thermal YM plasma at the end of inflation significantly affects the process of reheating.
We discuss the reheating after each inflation regime separately in the following.
For the sake of simplicity, we approximate the inflaton potential after inflation as Eq.~\eqref{eq:potential}.

In Secs.~\ref{sec:strw_reh}, \ref{sec:ww_reh}, and \ref{sec:c_reh}, we first consider the case of no glueball domination, whose condition is given in Eq.~\eqref{eq:prompt_glueball_decay}, and take $m_X = \bar{m}_X$ in Eq.~\eqref{eq:mX_max} for simplicity.
The results in this case are summarized in Fig.~\ref{fig:infreh}.
The case of the glueball domination is separately discussed in Sec.~\ref{sec:gluedom}, and the result is summarized in Fig.~\ref{fig:gD}.

\subsection{Reheating after strong warm inflation}
\label{sec:strw_reh}
Here we consider reheating after the strong warm inflation, \textit{i.e.}, $Q > 1$, which is realized for $\Lambda < \Lambda_\text{w-bdry}$ as given in \eqref{eq:wbdry}.

\vskip.3em
\noindent
\textbf{Dissipation of inflaton condensate.---}
Let us first discuss the evolution of the inflaton condensate governed by
\begin{equation}
  0 = \ddot \phi + \qty( 3 H + \Gamma_\phi ) \dot \phi + m_\phi^2 \phi.
\end{equation}
At the end of the strong warm inflation, the radiation energy density becomes comparable to that of inflaton [see Eq.~\eqref{eq:rho_rad_infend}], and the inequality, $\Gamma_\phi > m_\phi > H$, is fulfilled.
The inflaton equation of motion exhibits the over damping or damped oscillation, and the inflaton condensates dissipates its energy exponentially after inflation.
As the friction at a high temperature given in Eq.~\eqref{eq:thermal_fric} decreases in proportion to $T^3$, the dissipation is active until it becomes as small as $\Gamma_\phi \sim H$.
The suppression factor of the inflaton amplitude can be estimated as follows~\cite{BuenoSanchez:2010ygd}
\begin{equation}
  \phi \sim \phi_\text{end}\, \qty(\frac{a_\text{end}}{a })^{3/2} e^{- D}, \qquad D \equiv  \int^{T_\text{end}}_{T_\text{osc}} \frac{\dd T}{HT} \frac{m_\phi^2}{\Gamma_\phi (T)} +  \int^{T_\text{osc}}_{T_{\phi,\text{dec}}} \frac{\dd T}{H T} \frac{\Gamma_\phi(T)}{2},
\end{equation}
where $T_\text{end}$ represents the temperature at the end of strong warm inflation given in Eq.~\eqref{eq:Tend_str}, $T_\text{osc}$ is the temperature at the onset of inflaton oscillation $\Gamma_\phi (T) \sim m_\phi$, and $T_{\phi,\text{dec}}$ is the decoupling temperature of the inflaton condensate $\Gamma_\phi (T) \sim H$:
\begin{align}
  T_\text{end} &\sim \qty(\frac{\pi^2 g_\ast}{15})^{-1/4} V^{1/4}, \\
  T_\text{osc} &\sim \qty( \frac{d_\text{Ad} N^3}{4} )^{-1/3} \alpha^{-1} \qty( m_\phi \Lambda^2 )^{1/3}, \\
  T_{\phi,\text{dec}} &\sim \qty( \frac{d_\text{Ad} N^3}{4} )^{-1} \qty( \frac{\pi^2 g_\ast}{90} )^{1/2} \alpha^{-3} \qty( \frac{\Lambda^2}{\Mpl} ).
\end{align}
Given that the first integrand of $D$ is dominated by the lower end while the second one is dominated by the upper end, one may approximate the suppression factor by
\begin{equation}
  D \sim \frac{\Gamma_\phi (T_\text{osc})}{H_\text{osc}} 
  \sim \qty( \frac{d_\text{Ad} N^3}{4} )^{2/3} \qty( \frac{\pi^2 g_\ast}{90} )^{-1/2} \alpha^2 \, \frac{m_\phi^{1/3} \Mpl}{\Lambda^{4/3}}.
\end{equation}
Therefore, the inflaton condensate is expected to be dissipated away exponentially for
\begin{equation}
  \label{eq:nocond}
  \Lambda \ll
  \frac{d_\text{Ad}^{1/2} N^{3/2}}{2} \qty( \frac{\pi^2 g_\ast}{90} )^{-3/8} \alpha^{3/2} m_\phi^{1/4} \Mpl^{3/4}.
\end{equation}
This condition roughly coincides with the criteria for the strong warm inflation.

\vskip.3em
\noindent
\textbf{Inflaton-particle domination.---}
The efficient dissipation of the inflaton condensate does not immediately mean the completion of reheating.
This is basically because the production of inflaton particles occurs alongside the dissipation of inflaton condensate.
Indeed, within the range of $\alpha$ we consider, the friction rate is at most comparable to the inflaton particle production rate, \textit{i.e.,} $\overline{\Gamma}_{\delta \phi} \gtrsim \Gamma_\phi$.
Hence, the strong warm inflation, $\Gamma_\phi > 3 H$, automatically ensures that the inflaton particle is initially thermally populated due to $\overline{\Gamma}_{\delta \phi} \gtrsim \Gamma_\phi > H$.
As the production rate decreases in proportion to $T^3$, the inflaton particles eventually decouple from the thermal YM plasma at some point and behave as an independent population. When the temperature becomes comparable to the inflaton mass, $m_\phi \sim T$, the inflaton particles start to behave as pressureless matter, and its energy density with respect to the thermal YM plasma grows in proportion to the scale factor $a$. Finally, the inflaton particles may temporarily dominate the universe before eventually decaying perturbatively.

Let us first consider the case of $m_\phi > \Lambda_\text{conf}$ (see the non cross-hatched region in Fig.~\ref{fig:infend}).
In this case, the decay rate of inflaton particle is given in Eq.~\eqref{eq:vac_decay}, which occurs at $\overline{\Gamma}_{\delta \phi} \sim H$, \textit{i.e.,}
\begin{equation}
  \label{eq:pert_TR}
  T_{\text{R}} \sim \frac{d_\text{Ad}^{1/2}}{16 \pi^{3/2}}\, \qty( \frac{\pi^2 g_\ast}{90} )^{-1/4} \qty( \frac{\Mpl m_\phi}{\Lambda^2} )^{1/2} m_\phi.
\end{equation}
The decay rate is averaged over the Maxwellian distribution of inflaton particles with the weight of the boost factor, which leads to $\overline{\Gamma}_{\delta \phi} \sim \Gamma_\phi$ for $m_\phi \gg T$.
If this occurs after the inflaton-particle domination, the reheating is completed by the decay of inflaton particles and the additional dilution is expected from this process.
The inflaton-particle domination occurs if $1 < ( m_\phi / T_{\text{R}}) (\rho_{\delta \phi} / \rho_\text{rad} )_{T \sim m_\phi}$, which implies
\begin{equation}
  \label{eq:inflatonp-domination}
  \Lambda > 
  \frac{3^{1/2} 5^{1/4}}{2^{15/4} \pi^2}\, d_\text{Ad}^{1/2} \,g_\ast^{3/4}\, \qty( \Mpl m_\phi )^{1/2}.
\end{equation}

Now we move on to the case of $m_\phi < \Lambda_\text{conf}$ (see the cross-hatched region in Fig.~\ref{fig:infend}).
In this case, the decay rate of inflaton particle is no longer described by Eq.~\eqref{eq:vac_decay}, and it depends on a specific modeling of coupling between the YM sector and $\mathcal{L}_\text{others}$.
As a reference, we consider the case where the inflaton decay rate is given by Eqs.~\eqref{eq:prompt_glueball_decay_inflaton_decay} and \eqref{eq:mX_max}, leading to the reheating temperature of
\begin{equation}
  \label{eq:prompt_TR}
  T_{\text{R}}' \sim \qty( \frac{\pi^2 g_\ast}{90} )^{-1/4} \qty( \frac{\Mpl}{m_\phi} )^{1/2} \frac{\Lambda_\text{conf}^4}{\Lambda \bar{m}_X^2}
  \sim
  g_\ast^{-1} \qty( \frac{\Lambda_\text{conf}^5}{\Lambda^2 m_\phi} )^{1/2}.
\end{equation}
The inflaton-particle domination occurs if\footnote{
    The glueball dominantly decays into the inflaton particles for $m_X \gtrsim \Lambda$.
    As this further enhances the inflaton-particle domination, this condition can be relaxed by a few factors in some cases, strictly speaking.
    We neglect this effect because the inflaton particle domination is already inevitable in our case. See also the green-shaded region in Fig.~\ref{fig:infreh}.
}
\begin{equation}
  \Lambda > g_\ast \qty( \frac{\pi^2 g_\ast}{90} )^{-1/4} \qty( \frac{\Mpl}{m_\phi} )^{1/2} \frac{\Lambda_\text{conf}^4}{m_\phi \bar{m}_X^2}
  \sim \qty( \frac{\Lambda_\text{conf}^5}{m_\phi^3} )^{1/2}.
\end{equation}
As shown earlier in Eqs.~\eqref{eq:conf_end} and \eqref{eq:conf_end_c}, one may rewrite the confinement scale for a given $\Lambda$ and $\alpha_\text{end}$.
Inserting this, we obtain the condition for the inflaton-particle domination as a green-shaded region in the cross-hatched region in Fig.~\ref{fig:infreh}.
One may see that the inflaton-particle domination is inevitable in our case.

\subsection{Reheating after weak warm inflation}
\label{sec:ww_reh}
In this subsection, we discuss reheating after the weak warm inflation, $Q < 1$, 
which requires $\Lambda_\text{w-dbry} < \Lambda$. Additionally, depending on whether an initial thermal YM plasma exists, either $\Lambda < \Lambda_\text{w-th}$ or $\Lambda < \min [\Lambda_\text{w-th}, \Lambda_\text{warmed-up}]$ should be satisfied.

\vskip.3em
\noindent
\textbf{Inefficient dissipation.---}
Let us first consider the case where the preheating does not take place after the weak warm inflation.
This condition will be clarified shortly.
The radiation energy density at the end of inflation is smaller than that of inflaton, $\rho_\phi \gg \rho_\text{rad}$, for the weak warm inflation due to $Q < 1$ [see Eq.~\eqref{eq:rho_rad_infend}].
For each Hubble time, the amount of radiation generated from inflaton is $\Gamma_\phi \rho_\phi / H$.
At the end of warm inflation, the thermalization rate exceeds the Hubble parameter by definition.
Suppose that this component alone is thermalized. Then, the temperature is $T \propto H \propto a^{-3/2}$ decreasing faster than $T \propto a^{-1}$, which implies that the production is dominated right after inflation.
Hence, the temperature of radiation after the weak warm inflation can be estimated as
\begin{equation}
  \label{eq:Tweak_osc}
  T \sim \frac{\sqrt{3} d_\text{Ad} N^3}{16} \qty( \frac{\pi^2 g_\ast}{30} )^{-1} \alpha^3 \frac{\Mpl V^{1/2}}{\Lambda^2} \frac{a_\text{end}}{a}.
\end{equation}

An immediate consequence is that the inflaton condensate cannot be thermally dissipated away in this regime because the friction term never overcomes the Hubble parameter due to $Q \propto a^{-3/2}$.
The inflaton condensate behaves as pressureless matter and dominates the Universe until its decay rate becomes comparable to the Hubble parameter, $\Gamma_\phi \sim H$.
The reheating temperature is given by $T_\text{R}$ given in Eq.~\eqref{eq:pert_TR} as $\Gamma_\phi \sim \overline{\Gamma}_{\delta \phi}$ at $m_\phi \gg T$.

\vskip.3em
\noindent
\textbf{Preheating after weak warm inflation.---}
After the weak warm inflation, the inflaton oscillates its minimum with a period of $1/m_\phi$.
The instability scale is given by
\begin{equation}
  k_\text{inst} \equiv \frac{\Phi m_\phi}{2 \pi \Lambda},
\end{equation}
below which the modes would exhibit the tachyonic instability for each oscillation.\footnote{
This scale roughly coincides with $|\dot \phi|/(2 \pi \Lambda)$ during inflation for $\epsilon_V \sim 1$, as expected.
}
Here $\Phi$ represents the amplitude of inflaton oscillations, which decreases as $\Phi \propto a^{-3/2}$ due to the cosmic expansion.
Since the thermalization rate exceeds the Hubble parameter after the weak warm inflation, the magnetic screening mass, $m_\text{M} \simeq g^2 N T$, can be used as a criteria to have the strong instability, \textit{i.e.,} $k_\text{inst} > m_\text{M}$ and $k_\text{inst} \gg m_\phi$.
Note that, as the instability scale decreases faster than the temperature, an efficient tachyonic instability does not occur unless the instability scale is larger than the magnetic screening mass right after inflation.

\begin{figure}[t]
	\centering
  \includegraphics[height=0.25\linewidth]{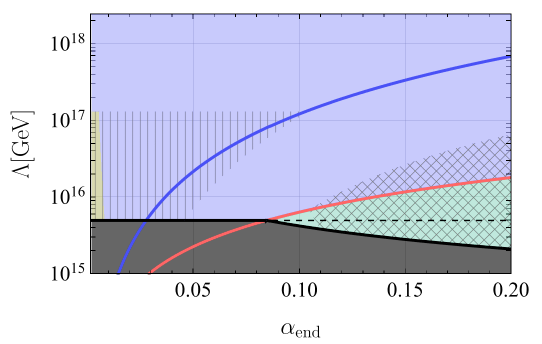}
  \hfill
  \includegraphics[height=0.25\linewidth]{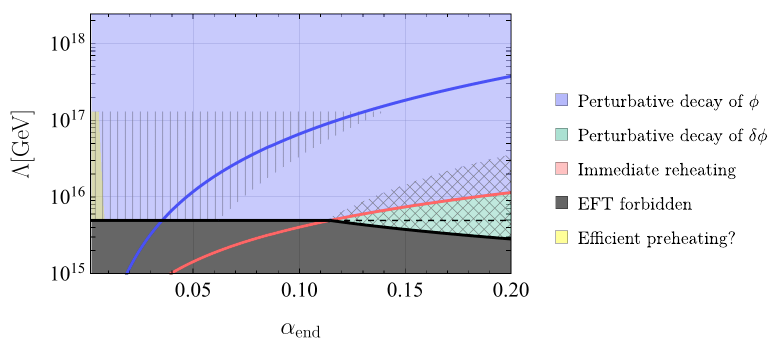}
	\caption{
    The completion of reheating after the axion inflation in each phase.
    The meaning of the solid/dashed lines and the parameters are the same as in Fig.~\ref{fig:infend}.
    The cross-hatched regions represent the parameter space where the inflaton decay rate is no longer described by Eq.~\eqref{eq:vac_decay}, where the inflaton decay rate strongly depends on a specific modeling of $\mathcal{L}_\text{others}$.
    We adopt Eqs.~\eqref{eq:prompt_glueball_decay_inflaton_decay} and \eqref{eq:mX_max} as a reference.
    We take $g_\ast = 30$ in the left panel, while $g_\ast = 100$ in the right panel.
    As mentioned in Sec.~\ref{sec:confinement}, we have taken the parameter $m_X$, which controls the coupling of the YM sector to $\mathcal{L}_\text{others}$, as $m_X = \bar{m}_X$ for simplicity.
    For this particular choice, there is no parameter space representing immediate reheating, \textit{i.e.}, a scenario in which reheating is completed right after inflation.
    For a smaller $m_X$, the parameters representing the immediate reheating appear in the cross-hatched region.
    }
    See Sec.~\ref{sec:strw_reh}.
    
	\label{fig:infreh}
\end{figure}

If the strong instability occurs, the amplitude of the gauge fields quickly becomes as large as $A_\text{NL}$, and the exponential growth is terminated.
The gauge fields are thermalized within one oscillation of inflaton, if the energy density exceeds a certain value:
\begin{equation}
  \label{eq:osc_th}
  \rho_\text{rad} \gtrsim \qty( \frac{\pi^2 g_\ast}{30} ) \qty( \frac{m_\phi}{c N^2 \alpha^2} )^4.
\end{equation}
The energy density of $A_\text{NL}$ is expected to be thermalized within one inflaton oscillation if
\begin{equation}
  \label{eq:prh_thermal}
  \Lambda \lesssim \frac{15^{1/4} c}{2^{5/2} \pi^{7/4}} N^{3/2} \alpha^{7/4} \Phi,
\end{equation}
which coincides with Eq.~\eqref{eq:nl_thermalization} for $\Phi = \sqrt{2} \Mpl$.
If this condition is fulfilled, the thermal YM plasma is generated for each oscillation and this production is eventually balanced by the cosmic expansion.
This consideration indicates that the following amount of radiation is generated for one cosmic time of $1/H$:
\begin{equation}
  \rho_\text{rad} \sim \rho_\text{rad}^\text{NL} \times \frac{m_\phi}{H}
  \sim
  \frac{3^{1/2}c}{2^{19/2} \pi^5} \frac{d_\text{Ad}}{N^2} \frac{\Mpl \Phi^3 }{\Lambda^4} \alpha^{-1} m_\phi^4 \propto a^{-9/2}.
\end{equation}
As it decreases faster than $a^{-4}$, the dominant production occurs right after inflation, whose amount is
\begin{equation}
  \label{eq:np_prod}
  \rho_\text{rad} \sim \frac{3^{1/2}c}{2^{15/2} \pi^5} \frac{d_\text{Ad}}{N^2} \frac{V_\text{end}^2}{\Lambda^4} \frac{\Mpl}{\Phi_\text{end}} \alpha^{-1} \times \qty(\frac{a_\text{end}}{a})^4.
\end{equation}
This contribution \eqref{eq:np_prod} is smaller than that generated via the thermal friction term in the warm inflation attractor \eqref{eq:Tweak_osc} if
\begin{equation}
  \Lambda \lesssim 2 \times 10^{18} \,\mathrm{GeV}\, \qty( \frac{d_\text{Ad}}{8} )^2 \qty( \frac{N}{3} )^{7/2} \qty( \frac{\Phi_\text{end}}{\Mpl} )^{1/4} \qty( \frac{\alpha}{0.1} )^{13/4},
\end{equation}
which holds for the parameters of our interest. See also Fig.~\ref{fig:infreh}.
Therefore, the inflaton condensate continues to dominate the Universe as in the previous case, and reheats the Universe at $T_\text{R}$.

\subsection{Reheating after cold inflation}
\label{sec:c_reh}
In this subsection, we discuss reheating after the cold inflation, which is realized in $\Lambda > \Lambda_\text{w-th}$ or $\Lambda > \min [\Lambda_\text{w-th}, \Lambda_\text{warmed-up}]$ depending on the existence of initial thermal YM plasma.

\vskip.3em
\noindent
\textbf{No preheating.---}
After inflation, the instability scale decreases in proportion to $a^{-3/2}$ due to the cosmic expansion.
Hence, if the instability scale is not large enough immediately after inflation, the tachyonic instability never occurs later.
Motivated by the numerical lattice simulations in axion inflaton coupled to $\mathrm{U}(1)$ gauge theory~\cite{Adshead:2015pva,Cuissa:2018oiw,Figueroa:2024rkr}, we adopt $k_\text{inst} / m_\phi > 3$ for the occurrence of the strong instability.
Such an efficient production does not occur for
\begin{equation}
  \label{eq:noprh}
  \Lambda \gtrsim 10^{17}\,\mathrm{GeV} \times \qty( \frac{\Phi_\text{end}}{\Mpl} ).
\end{equation}
In this case, the oscillation of the inflaton condensate dominates the Universe after inflation, and the reheating is completed by the perturbative decay at $T_\text{R}$.

\vskip.3em
\noindent
\textbf{Preheating after inflation.---}
If Eq.~\eqref{eq:noprh} is violated, the tachyonic instability occurs after inflation.
As discussed in the previous subsection, the amplitude of the gauge fields immediately becomes as large as $A_\text{NL}$ within one oscillation.

Let us first consider the case where the gauge fields are quickly thermalized for each inflaton oscillation, which is realized for the cutoff scale smaller than a certain value given in Eq.~\eqref{eq:prh_thermal}.
The amount of radiation generated in this case is given by Eq.~\eqref{eq:np_prod}, which is smaller than the energy density of inflaton condensate for
\begin{equation}
  \Lambda \gtrsim 7.3 \times 10^{14} \,\mathrm{GeV}\, \qty( \frac{d_\text{Ad}}{3} )^{1/4} \qty( \frac{N}{3} )^{-1/2} \qty( \frac{\Phi}{\Mpl} )^{-1/4} \qty( \frac{\alpha}{0.1} )^{-1/4}.
\end{equation}
In the parameters of our interest, this condition is automatically fulfilled (see Fig.~\ref{fig:infreh}).
Thus, the inflaton condensate continues to dominate the Universe and the reheating is completed by the perturbative decay at $T_\text{R}$.

On the other hand, if the thermalization does not occur for each inflaton oscillation, one has to deal with non-equilibrium physics.
Once the amplitude of the gauge fields becomes as large as $A_\text{NL}$, the exponential growth is terminated, and the system is governed by the turbulent phenomena, which typically obeys a certain scaling behavior.
A proper treatment in this case is to perform the numerical lattice simulation that is beyond the scope of this paper.
Nevertheless, restricting our interest to the completion of reheating, we can derive the following criteria whether the reheating is completed by the perturbative decay.

Let us be agnostic about a possible non-thermal energy conversion from the inflaton to radiation, and ask a question whether a significant amount of inflaton energy density can be extracted.
If the energy density of radiation somehow becomes as large as the right-hand side of Eq.~\eqref{eq:osc_th}, the gauge fields get thermalized within one inflaton oscillation.
This implies that the energy density of radiation cannot exceed the following value
\begin{equation}
  \rho_\text{rad} < \qty( \frac{\pi^2 g_\ast}{30} ) \qty( \frac{m_\phi}{c N^2 \alpha^2} )^4 \times \frac{m_\phi}{H}
  \lesssim
  3 \,\qty( \frac{d_\text{Ad}}{8} ) \qty( \frac{N}{3} )^{-8} \qty( \frac{\alpha}{0.01} )^{-8} \qty( \frac{\Lambda}{10^{15}\,\mathrm{GeV}} )^{-1} \times \rho_\text{inf}.
\end{equation}
Here we insert a possible maximum value of the right-hand side, where the Hubble parameter is evaluated at the end of the efficient production, \textit{i.e.,} $k_\text{inst} / m_\phi \sim 3$.
Nevertheless, the right-hand side is always smaller than the inflaton energy density in most parameter of our interest (see the yellow shaded region in Fig.~\ref{fig:infreh}), and hence the inflaton condensate continues to dominate the Universe.
Therefore, even in this case, we expect that the reheating is completed by the perturbative decay at $T_\text{R}$.

\subsection{Glueball domination}
\label{sec:gluedom}

\begin{figure}[t]
	\centering
  \includegraphics[height=0.25\linewidth]{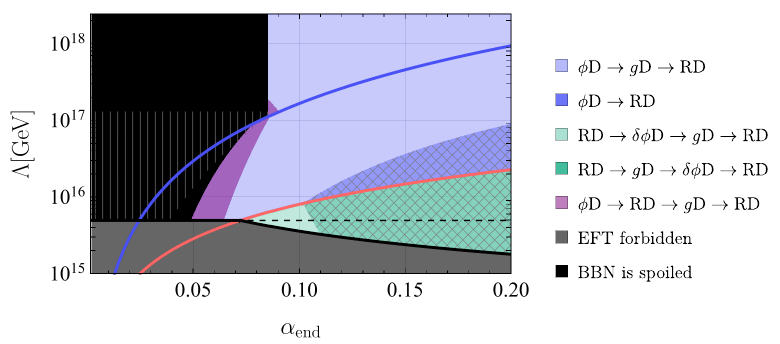}
	\caption{
    The same figure as in Fig.~\ref{fig:infreh} except for the glueball domination.
    The black region is excluded by the requirement of the glueball decay before the BBN epoch.
    The glueball, inflaton condensate, and inflaton particle domination are denoted by $g \text{D}$, $\phi \text{D}$, and $\delta \phi \text{D}$ respectively.
    }
	\label{fig:gD}
\end{figure}

So far we have assumed that the glueball never dominates the Universe by requiring Eq.~\eqref{eq:prompt_glueball_decay}.
Once we relax this assumption, the process of reheating is significantly affected by the glueball production and the subsequent dilution by its decay.
The comprehensive discussion of all the possible cosmological evolution is beyond the scope of this paper, and we leave it for future work.
Instead, we focus on the extreme case where the glueball domination lasts for a significant period by fixing $m_X = \Mpl$.
As discussed in Sec.~\ref{sec:confinement}, the glueball decays into the inflaton particles as long as the decay channel is open, \textit{i.e.,} $\Lambda_\text{conf} > m_\phi$.
Since $m_X = \Mpl > \Lambda$ is fulfilled in the cross-hatched region of Fig.~\ref{fig:gD}, \textit{i.e.,} $\Lambda_\text{conf} > m_\phi$, the glueball dominantly decays into the inflaton particles in the cross-hatched region in this case.

If the inflaton mass is larger than the confinement scale, \textit{i.e.,} $m_\phi > \Lambda_\text{conf}$ (non-cross-hatched region in Fig.~\ref{fig:gD}), the glueball decay is governed by Eq.~\eqref{eq:glueball_decayrate_other}.
By comparing the glueball decay rate of Eq.~\eqref{eq:glueball_decayrate_other} with the inflaton decay rate given in Eq.~\eqref{eq:vac_decay}, one may show that the glueball decay temperature is always smaller than the inflaton decay temperature $T_\text{glueball} < T_\text{R}$ because
\begin{equation}
  \frac{\Gamma_\text{glueball}}{\Gamma_\phi} \sim \frac{\Lambda^2 \Lambda_\text{conf}^5}{m_\phi^3 \Mpl^4}
  < \frac{\Lambda^2 m_\phi^2}{\Mpl^4},
\end{equation}
where the right-hand side is always much less than unity in the parameters of our interest.
Hence, in this case, the Universe is once dominated by the glueballs, and then the glueball decay completes the reheating.
In order not to spoil the success of the Big Bang nucleosynthesis (BBN), the glueball should decay before the BBN epoch.
This consideration puts a lower bound on the confinement scale as
\begin{equation}
  \label{eq:Tglueball}
  1\,\text{MeV} \lesssim T_\text{glueball} \sim 3\, \text{MeV} \times \qty( \frac{\Lambda_\text{conf}}{10^{10}\,\mathrm{GeV}} )^{5/2},
\end{equation}
which excludes the black region in Fig.~\ref{fig:gD}.

On the other hand, for $m_\phi < \Lambda_\text{conf}$ (cross-hatched region in Fig.~\ref{fig:gD}), the glueball dominantly decays into the inflaton particles, whose decay rate is given by Eq.\eqref{eq:glueball_decayrateinto_inflaton}.
By comparing this rate with the inflaton decay rate for $m_\phi < \Lambda_\text{conf}$ given in Eq.~\eqref{eq:prompt_glueball_decay_inflaton_decay}, we obtain
\begin{equation}
  \frac{\Gamma_\phi}{\Gamma_\text{glueball}} \sim \frac{\Lambda_\text{conf}^3 \Lambda^2}{\Mpl^4 m_\phi}
  \lesssim \frac{\Lambda_\text{conf} \Lambda^2}{\Mpl^3},
\end{equation}
where the right-hand side is always much less than unity in the parameters of our interest.
Here we have used that the confinement scale is smaller than the inflaton potential in our parameters, $m_\phi \Mpl > \Lambda_\text{conf}^2$.
Hence, the inflaton decay occurs after the glueball decay, \textit{i.e.,} $T_\text{R} < T_\text{glueball}$, and the reheating is completed by the inflaton decay.
The reheating temperature is estimated as
\begin{equation}
  T_\text{R} \sim 10^3 \, \text{GeV}\, \qty( \frac{\Lambda_\text{conf}}{10^{13}\,\text{GeV}} )^4 \qty( \frac{\Lambda}{10^{15}\,\text{GeV}} )^{-1}.
\end{equation}
One may confirm that this is always larger than the BBN temperature $\sim 1\,\text{MeV}$ in the cross-hatched region of Fig.~\ref{fig:gD}.

\vskip.3em
\noindent
\textbf{After the weak warm or cold inflation.---}
In these cases, the inflaton condensate remains after inflation and first dominates the Universe.
The subsequent cosmological evolution depends on whether the inflaton mass is larger or smaller than the confinement scale.

For $m_\phi > \Lambda_\text{conf}$, the inflaton condensate decays perturbatively at $T_\text{R}$ into the YM sector.
In the case of $T_\text{R} > \Lambda_\text{conf}$, the YM sector remains in the deconfined phase at $T_\text{R}$, and the Universe is dominated by radiation after the inflaton decay.
After the confinement, the glueball starts to dominate the Universe until it decays at $T_\text{glueball}$. 
See the purple region in Fig.~\ref{fig:gD}.
On the other hand, in the case of $T_\text{R} < \Lambda_\text{conf}$, the confinement occurs before the inflaton decay.
The inflaton dominantly decays into glueballs at $T_\text{R}$, and the glueball dominates the Universe until it decays at $T_\text{glueball}$. See the light blue region in Fig.~\ref{fig:gD}.

For $m_\phi < \Lambda_\text{conf}$, the inflaton condensate perturbatively decays into the light sector of $\mathcal{L}_\text{others}$ via Eq.~\eqref{eq:prompt_glueball_decay_inflaton_decay}.
Owing to $T_\text{R} < T_{\text{glueball}}$, the glueballs decay before the inflaton decay, primarily into inflaton particles.
The inflaton condensate remains the dominant component until its decay.
Hence, the reheating is completed by the perturbative decay of the inflaton condensate at $T_\text{R}$.
See the blue region in Fig.~\ref{fig:gD}.

\vskip.3em
\noindent
\textbf{After the strong warm inflation.---}
Since the confinement scale is smaller than the inflation scale at the end of inflation, the YM sector still remains in the deconfined phase, and the Universe is once dominated by radiation after inflation.
The subsequent evolution again depends on whether the inflaton mass is larger or smaller than the confinement scale.

For $m_\phi > \Lambda_\text{conf}$, the thermally produced inflaton particles eventually dominate the Universe once they become non-relativistic. These particles decay into glueballs at $T_\text{R}$, leading to a glueball-dominated era. Later, the glueballs decay perturbatively at $T_\text{glueball}$, after which radiation again becomes dominant. 
See the light green region in Fig.~\ref{fig:gD}.

For $m_\phi < \Lambda_\text{conf}$, the glueball dominates the Universe after the confinement transition.
The glueball decays at $T_\text{glueball}$ dominantly into the inflaton particles, and the inflaton particles dominate the Universe until they decay perturbatively at $T_\text{R}$.
See the green region in Fig.~\ref{fig:gD}.

\section{Summary and Discussion}
\label{sec:sumdis}

In this paper, we have studied the reheating process after axion inflation coupled to the non-Abelian gauge theory via the CS coupling.
We have shown that the reheating is completed by the perturbative decay of either the inflaton condensate, the inflaton particle or the glueball in a broad parameter space.
A naive expectation was that the gauge field production via the tachyonic instability dissipates a significant amount of the inflaton condensate, completing the reheating.
However, the produced gauge fields get thermalized by their own self-interactions, and the production of the inflaton particles or the glueballs from the thermal bath can lead to the subsequent domination of the Universe by them.
In addition, the non-Abelian nature suppresses the resonant production owing to the self-interactions, resulting in the remnant of the inflaton condensate even after the tachyonic instability in some parameter spaces.
Owing to these effects, the Universe is dominated by the inflaton condensate, the inflaton particle, or the glueball, and in most parameter space of our interest, their perturbative decay completes the reheating (See Figs.~\ref{fig:cls}, \ref{fig:infreh} and \ref{fig:gD}).

The phenomenological implications of the early matter domination by the inflaton or the glueball after axion inflation are diverse.
For instance, the inflaton particle domination after the strong warm inflation could alter the conventional predictions, which are based on the radiation domination immediately after the strong warm inflation.
The subsequent perturbative decay not only results in entropy dilution~\cite{BuenoSanchez:2010ygd}, but also provides alternative mechanisms for dark matter production~\cite{Harigaya:2014waa,Harigaya:2019tzu}, baryon asymmetry generation~\cite{Hamada:2015xva,Hamada:2018epb,Blazek:2024efd}, and gravitational wave production~\cite{Ema:2015dka,Nakayama:2018ptw}, for instance.

The primary purpose of this paper is to illustrate the intricate dynamics of the reheating process following axion inflation coupled to a non-Abelian gauge theory, and to underscore the critical role of massive particles, \textit{i.e.}, the inflaton particles and glueballs, in this transition.
For this reason, our analysis is based on a simplified inflaton model, where the inflaton potential is dominated by Eq.~\eqref{eq:potential} after inflation and the inflation end is assumed to be determined by the first slow-roll condition.
Although this is a reasonable assumption for a simple high-scale inflation, the inflaton potential can be more complex in general, and the inflation end can be determined by the second slow-roll condition.
We have also assumed that the non-Abelian gauge theory is described by the (almost) pure YM theory in the dark sector.
In reality, the dark sector can involve charged matter fields, or the gauge fields can be related to the SM gauge group.
We leave these generalizations for future work.

One of the key features of this setup is the self-interactions of the non-Abelian gauge fields, which thermalizes the YM plasma, allows the inflaton particle production via Eq.~\eqref{eq:inflaton_production}, and suppresses the tachyonic instability.
For a smaller gauge coupling, the effect of self-interactions becomes less significant, and we expect that the system becomes more close to the pure Abelian gauge theory.
There the system is highly nonthermal which calls for a lattice simulation.
This picture motivates us to anticipate that the reheating process after the axion inflation coupled to the Abelian gauge theory can be significantly modified once the charged matter fields are introduced~\cite{Domcke:2018eki,Domcke:2018gfr,Domcke:2019qmm,Gorbar:2021rlt,Fujita:2022fwc,vonEckardstein:2024tix}.
The interactions of the Abelian gauge fields with the charged matter fields could lead to the similar effects to the non-Abelian case, such as the thermalization, the inflaton particle production, and the suppression of the tachyonic instability.
The detailed analysis including lattice simulations in this case is worth investigating in the future.

\section*{Acknowledgement}
We thank Mikko Laine and Kazunori Kohri for useful discussions. We were supported by JSPS KAKENHI Grant No.\ JP23K03424 (T.\,F.\,) and JP22K14044 (K.\,M.\,).

\appendix

\small
\bibliographystyle{utphys}
\bibliography{ref}

\end{document}